\documentstyle[epsfig,psfig]{mn} 

\title[An electrically powered binary star?]
{ An electrically powered binary star? }

\author[K. Wu et al.]{Kinwah Wu$^{1,2}$, Mark Cropper$^{2}$, 
     Gavin Ramsay$^{2}$ and Kazuhiro Sekiguchi$^{3}$ \\ 
$^{1}$Research Centre for Theoretical Astrophysics, 
       School of Physics A28,
       University of Sydney, NSW 2006, Australia \\
$^{2}$Mullard Space Science Laboratory, University College London, 
       Holmbury St.~Mary, Dorking, Surrey, RH5~6NT \\    
$^{3}$National Astronomical Observatory of Japan, 
       650 Nth A'ohoku Place, Hilo, HI 96720, USA  }  

\date{Received: }

\begin{document} 
 
\def\Mdot{\hbox{$\dot M$}}
\def\Msun{\hbox{M$_\odot$}}
\def\Rsun{\hbox{$R_\odot$}} 
\outer\def\gtae {$\buildrel {\lower3pt\hbox{$>$}} \over 
       {\lower2pt\hbox{$\sim$}}$}
\outer\def\ltae {$\buildrel {\lower3pt\hbox{$<$}} \over
       {\lower2pt\hbox{$\sim$}}$}
\def\rchi{{${\chi}_{\nu}^{2}$}}   
\def\ddt{{d \over {dt}}} 

\maketitle

\begin{abstract} 

We propose a model for stellar binary systems 
   consisting of a magnetic and a non-magnetic white-dwarf pair 
   which is powered principally by electrical energy. 
In our model the luminosity is caused by resistive heating 
   of the stellar atmospheres  
   due to induced currents driven within the binary.  
This process is reminiscent of the Jupiter-Io system, 
   but greatly increased in power 
   because of the larger companion 
   and stronger magnetic field of the primary.  
Electrical power is an alternative stellar luminosity source,
   following on from nuclear fusion and accretion.   
We find that this source of heating is sufficient 
   to account for the observed X-ray luminosity 
   of the 9.5-min binary RX~J1914+24,     
   and provides an explanation for its puzzling characteristics.  

\end{abstract}


\section{Introduction} 

It has been observed directly by the {\it Hubble Space Telescope} that 
   the movement of Io through Jupiter's magnetic field 
   causes heating in the Jovian atmosphere (Clarke et al.\ 1996).  
This is because a conducting body transversing a magnetic field 
   produces an induced electric field. 
When the circuit is closed, a current will be set up, 
   resulting in resistive dissipation. 
The Jupiter-Io system therefore operates as a unipolar inductor  
   (Paddington \& Drake 1968; Goldreich \& Lynden-Bell 1969). 
Another potential cosmic unipolar inductor could be a planet orbiting
   around a magnetic white dwarf (Li, Ferrario \& Wickramasinghe 1998).  
These systems have a similar configuration, 
   with the differences being their orbital period and separation, 
   the masses and radii of the two components,  
   and the magnetic moment of the magnetic object.   

We propose that binary stars 
   consisting of a magnetic and a non-magnetic white dwarf 
   can also be cosmic unipolar inductors (Fig.~1).  
Close binaries of this type can have short periods 
   and secondaries larger than planet-sized bodies.  
Provided that the spin of the magnetic white dwarf 
   and the orbital rotation 
   are not synchronised  
   (so that the secondary is in motion relative to the magnetic field) 
   and that the density of the plasma between the white dwarfs 
   is high enough, 
   unipolar induction will operate efficiently. 

Gravitational waves carry away the orbital angular momentum efficiently 
   from a short-period ($<$2~hr) binary system 
   but not the stellar spin momenta directly.  
In a magnetic and non-magnetic white-dwarf pair 
   with only the non-magnetic star tidally locked, 
   the magnetic star will be spun up retrogradely 
   in the orbital rest frame 
   as the binary orbit shrinks. 
Alternatively, the magnetic white dwarf could be spun up by accretion  
   in a previous epoch in which mass transfer occurred,  
   so that it has a spin faster than the orbital rotation.    
Asynchronous rotation can therefore occur   
   as long as a stellar component is not tidally or otherwise locked. 
  
The dissipative power of a white dwarf--planet pair  
   with an orbital period $P = 10$~hr  
   is estimated to be $\sim 10^{29}{\rm erg~s}^{-1}$ 
   (Li, Ferrario \& Wickramasinghe 1998). 
The output of a binary white-dwarf pair with a period of tens of minutes 
   will be significantly higher  
   and should have detectable observational consequences.    

In this paper, a simple model for a unipolar inductor   
   consisting of a magnetic and a non-magnetic white-dwarf pair 
   in a close orbit is presented.  
The basic features of these binaries and their observational properties 
   are discussed. 
We propose that the short-period soft X-ray source RX~J1914+24  
   (Cropper et al.\ 1998; Ramsay et al.\ 2000a) 
   is a candidate unipolar inductor      
   consisting of a magnetic and non-magnetic white-dwarf pair.      

\section{White-dwarf pairs as cosmic unipolar inductors}   

\begin{figure}
\begin{center}  
\leavevmode
\setlength{\unitlength}{1cm}
\begin{picture}(8,6)
\put(-1,-1.5){\includegraphics{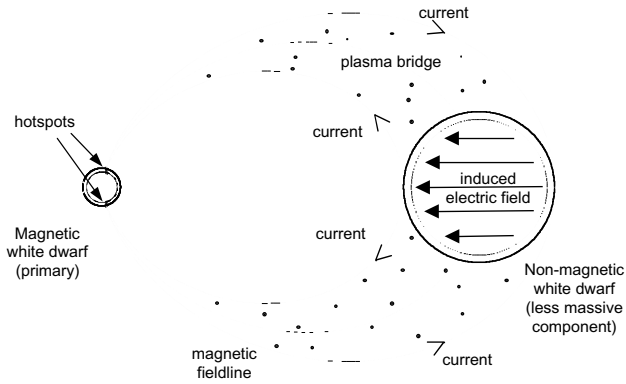}} 
\end{picture}
\end{center}
\caption{A schematic illustration of a unipolar inductor 
    consisting of a magnetic and non-magnetic white-dwarf pair 
    in a close binary orbit. }
\label{finding} 
\end{figure}

When a non-magnetic conductor of linear size $R$     
   transversing a magnetic field $\vec B$  
   with a velocity $\vec v$,  
   the induced e.m.f.\  across the conductor  is     
   $\Phi  \sim   R | {\vec E} |$, 
   where ${\vec E} = ({\vec v} \times {\vec B})/c$  
   ($c$ is the speed of light).   
Thus, the corresponding e.m.f.\ across a non-magnetic white dwarf 
   in orbit with a magnetic white dwarf is   
\begin{eqnarray}  
   \Phi  & \approx & {{2\pi} \over c} \biggl(  
      {{\mu_1 R_2} \over {a^2 P}} \biggr) (1 - \alpha) \nonumber \\ 
    &  = &  \biggl({{\mu_1 R_2} \over c} \biggr)  
        \biggl({{2\pi} \over P}  \biggr)^{7/3} 
       (1 - \alpha) \big[G M_1 (1+q) \big]^{-2/3} \  , 
\end{eqnarray}   
   where $G$ is the gravitational constant, 
   $q$ ($\equiv M_2/M_1$) is the ratio  
   of the non-magnetic to the magnetic white dwarf mass, 
   $R_2$ and $R_1$ are the radii  
   of the non-magnetic and the magnetic white dwarf respectively,      
   $\mu_1$ is the magnetic moment of the magnetic white dwarf. 
The degree of asynchronism $\alpha$ 
   is defined as the ratio 
   of the spin angular speed of the magnetic white dwarf $\omega_1$   
   to the orbital angular speed  $\omega_o$ ($=2\pi/P$). 
(Here and elsewhere, we consider that 
   the anti-clockwise direction is positive.)  
We show in Figure 2 the induced e.m.f.\ for different system parameters.  
The induced e.m.f.\ depends strongly on the orbital period, 
   the degree of spin-orbit synchronism,  
   and the mass (radius) of the non-magnetic white dwarf.  
The dependence of the mass of the magnetic white dwarf 
   is weaker.  

\begin{figure} 
\vspace*{0.2cm}  
\begin{center}
  \epsfxsize=8cm \epsfbox{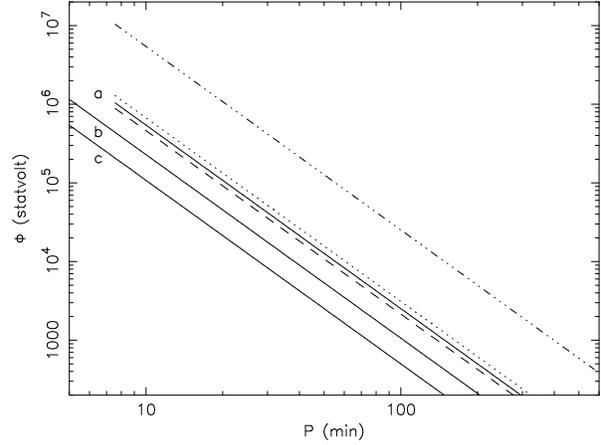}  
\end{center}
\caption{The induced e.m.f.\  
      across the non-magnetic secondary white dwarf 
      as a function of orbital period 
      for binaries with various parameters.  
   Solid lines a, b and c correspond to systems 
      with a primary magnetic white dwarf of 1.0~M$_\odot$ 
      with a spin 1 part of 1000 deviated 
      from synchronous rotation with the orbit.   
   The masses of the secondaries in these systems 
      are 0.1, 0,5 and 1.0~M$_\odot$ respectively.     
   The dotted line corresponds to systems 
      consisting of a 0.1-M$_\odot$ non-magnetic white dwarf, 
      and a 0.7-M$_\odot$ magnetic white dwarf 
      with 1 part of 1000 off spin-orbit synchronism; 
      and the dashed line corresponds to similar systems 
      but with a 1.3-M$_\odot$ primary magnetic white dwarf. 
   Systems with magnetic and non-magnetic white-dwarf pairs 
      of mass 1.0 and 0.5~M$_\odot$ 
      and spin-orbital asynchronism of 1 part in 100 
      for the magnetic white dwarf is represented 
      by the dot-dashed line. 
   The magnetic moments of the systems 
      are fixed to be $10^{32}$~G~cm$^3$.    }
\end{figure}    
 
If the space between the white dwarfs is filled with plasma, 
   the induced e.m.f.\ will drive currents 
   along the magnetic field lines connecting the two white dwarfs.  
Because there is substantial resistance in the white-dwarf atmosphere,   
   the white dwarfs act as the dissipative components  
   in this electrical circuit.   
The total electrical power dissipation in the two white dwarfs is  
\begin{eqnarray}   
   W &  = & I^2 ({\cal R}_1+{\cal R}_2) \nonumber \\ 
     &  = & \Phi^2/({\cal R}_1+ {\cal R}_1) \ ,    
\end{eqnarray} 
   where $I$ is the total current,   
   and ${\cal R}_1$ and ${\cal R}_2$ 
   are the effective resistance 
   of the magnetic and the non-magnetic white dwarfs respectively. 
An object with a length $L$ and an area $A$ 
   has a resistance ${\cal R} = L /A\sigma$  
   (where $\sigma$ is conductivity). 
Therefore, the ratio of the effective resistances 
   of the white dwarfs is    
\begin{eqnarray} 
   {{\cal R}_1} \over {{\cal R}_2} & \sim & 
      \biggl({{\sigma_2}\over {\sigma_1}} \biggr)
       \biggl({{R_2^2}\over {fR_1^2}}\biggr)
       \biggl({{\Delta h_1}\over {\Delta h_2}}\biggr)   \ ,  
\end{eqnarray}  
   where  $\sigma_1$ and $\sigma_2$ are the conductivity 
   of the magnetic and the non-magnetic white dwarf respectively,  
   $\Delta h_1$ and $\Delta h_2$ are the thickness 
   of the dissipative surface layers of the white dwarfs, and 
   $f$ is the fractional effective area of the magnetic poles (hot spots) 
   on the surface of the magnetic white dwarf.    
As $f \ll 1$ (see Appendix A), 
   the effective resistance of the magnetic white dwarf 
   is significantly larger than the non-magnetic white dwarf.   

As the currents pass through both white dwarfs,   
   the ratio of the power dissipation in the magnetic primary 
   to that of the non-magnetic secondary is 
   ${{W_1}/{W_2}} =  {{\cal R}_1} / {{\cal R}_2}$.  
Taking account of the geometry of the current loops,   
   we obtain 
\begin{eqnarray} 
    {{W_1} \over {W_2}}&  \approx & \beta 
      \biggl({{\sigma_2} \over {\sigma_1}} \biggr)   
      \biggl({{R_2} \over {\Delta R_2}} \biggr)     
      \biggl[{{G(M_1+M_2)} \over {R_1^3}}   
           \biggl({P \over {2\pi}} \biggr)^2 \biggr]^{1/2}  \ ; \\    
   {\cal R}_1 & \approx & {1\over {2 \sigma_1}} 
                    \biggl({{H}\over {\Delta d}}\biggr) 
                    \biggl({a \over R_1}\biggr)^{3/2} 
                    {{{\cal J}(e)} \over R_2} \ ;  \\ 
   {\cal R}_2 & \approx & {4\over {\pi \sigma_2}}  
                      \biggl({{\Delta R_2}\over {R_2^2}}\biggr)    
\end{eqnarray}  
  (see Appendices A and B for details), 
   where $\Delta R_2$ is the thickness of the secondary's atmosphere 
   and $\beta$ is a structure factor of the order of unity.  
The factor ${\cal J}(e)$ depends of the radii of the white dwarfs 
   relative to the orbital separation. 
Its value is of the order of unity for white-dwarf pairs 
   with orbital periods less than an hour.   

The conductivity of plasma of an electron temperature $T_{\rm e}$ 
   is given by 
\begin{eqnarray} 
   \sigma & = & \gamma \biggl({{2^{5/2}}\over {\pi^{3/2}}} \biggr)
        {{(kT_{\rm e})^{3/2}}\over{m_{\rm e}^{1/2}Ze^2 \ln \Lambda}} \ , 
\end{eqnarray}    
   (Spitzer \& H$\ddot {\rm a}$rm 1953)  
   where $k$ is the Boltzmann constant, 
   $T_{\rm e}$ is the electron temperature, 
   $m_{\rm e}$ is the electron mass,  
   $e$ is the electron charge,  
   $Z$ is the ion charge number, 
   and $\ln \Lambda$ is the Coulomb logarithm. 
The factor $\gamma$ depends on the ion charge number $Z$, 
   which has values between 0.6 ($Z=1$) and 1 ($Z\rightarrow \infty$)  
   (see Alfv$\acute {\rm e}$n \& F$\ddot {\rm a}$lthammar 1963).    
For a white-dwarf atmosphere with $T_{\rm e} \sim 10^5$~K, 
   the conductivity $\sigma \sim 10^{13}-10^{14}$~esu. 
Since the conductivities of the atmospheres 
   of white dwarfs are similar to each other,   
   the majority of the electrical power will be dissipated 
   in small regions at the footpoints 
   of the current-carrying field lines
   on the surface of the magnetic white dwarf.  

\section{Energy and angular-momentum conservation}   

\subsection{Power dissipation} 

\begin{figure} 
\vspace*{0.2cm}  
\begin{center}
  \epsfxsize=8cm \epsfbox{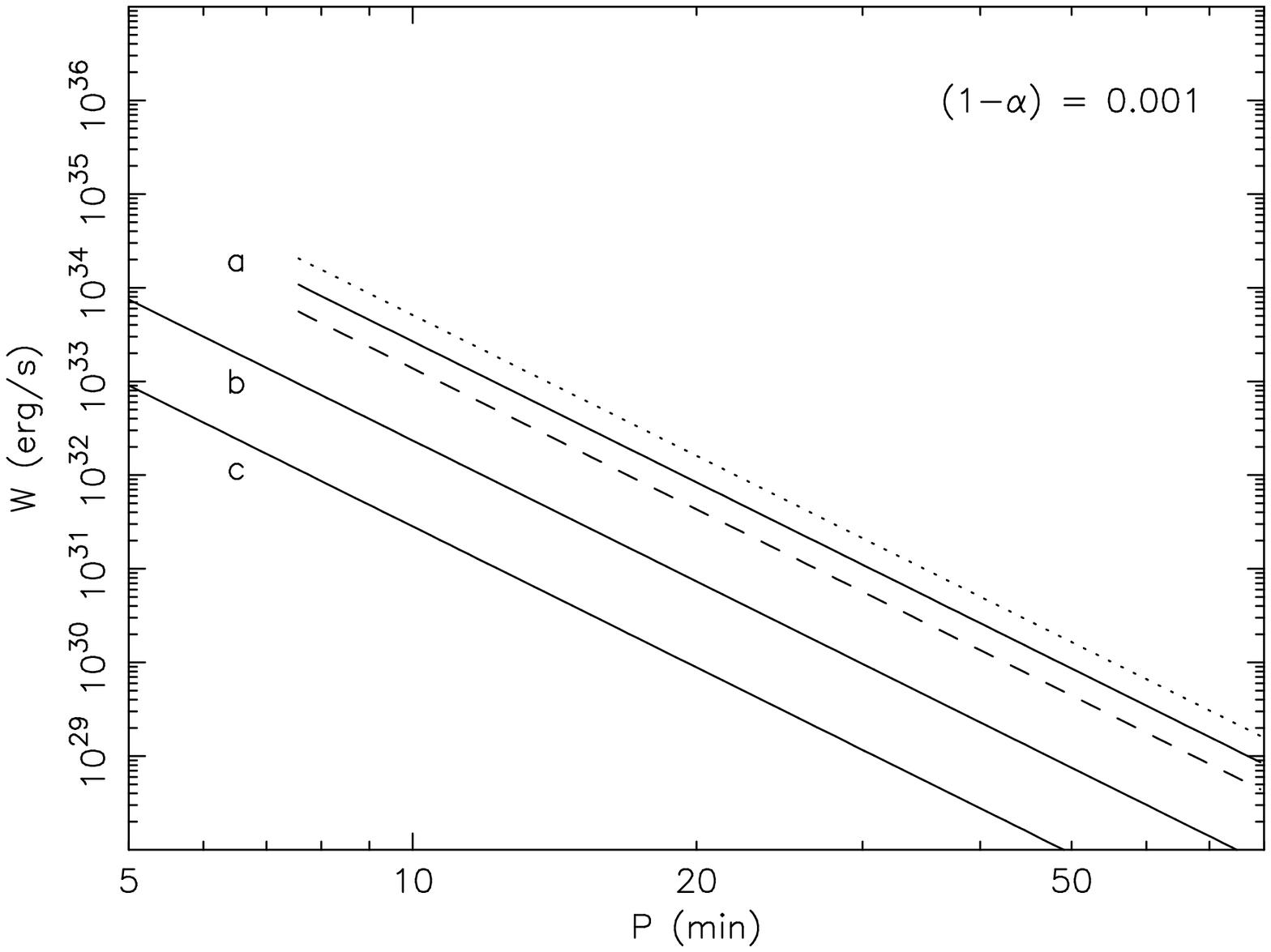}  \\  
   \epsfxsize=8cm \epsfbox{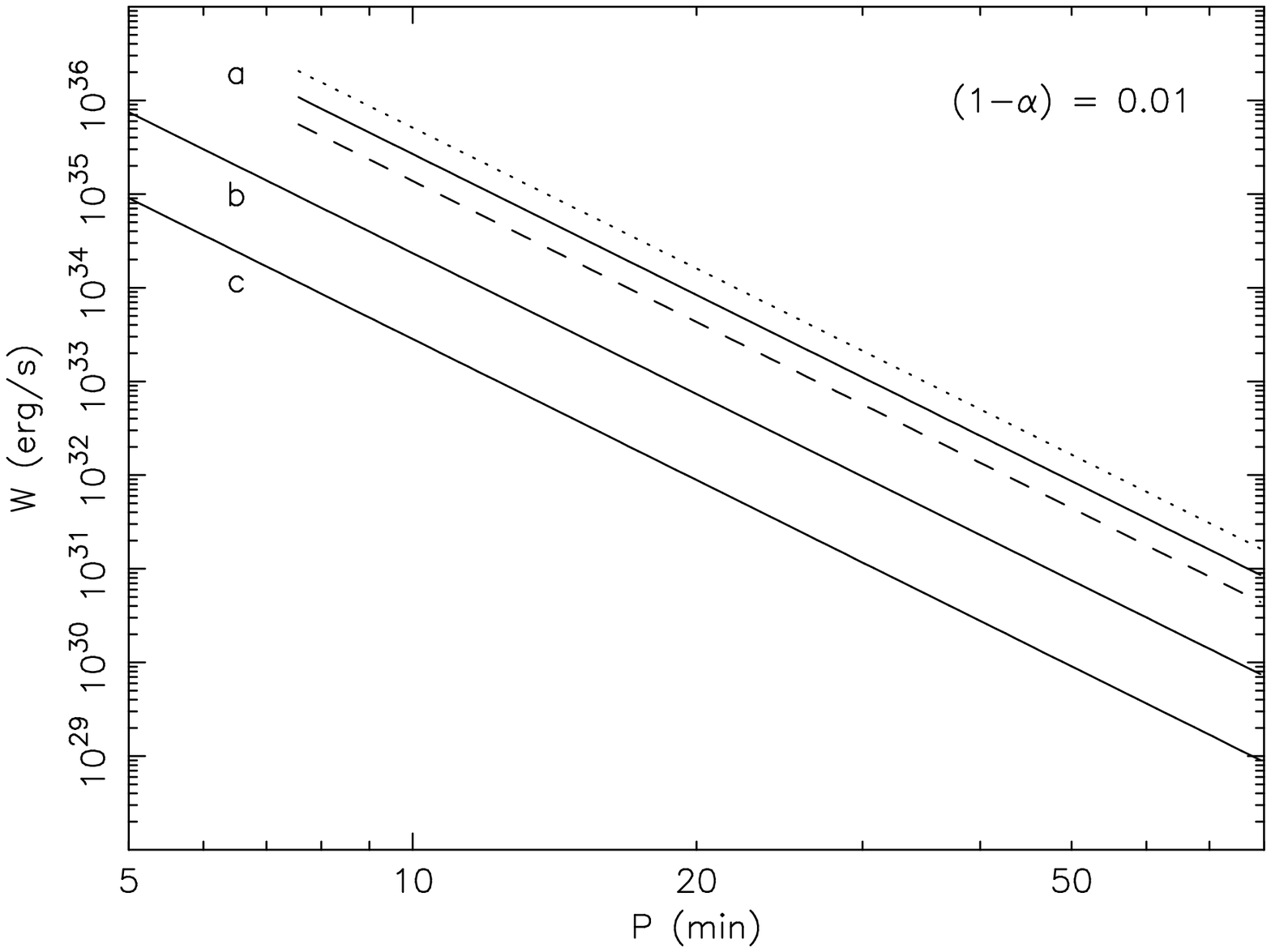}  
\end{center}
\caption{The total electrical power dissipation   
      as a function of orbital period. 
    The degree of deviation from spin-orbit asynchronism 
      $(1-\alpha)$ is 1 part in 1000 in the top panel and 
      1 part in 100 in the bottom panel. 
    As in Fig.~2 the solid lines a, b and c correspond 
      to systems with a non-magnetic white-dwarf secondary 
      of 0.1, 0.5 and 1.0~M$_\odot$. 
    The mass of the magnetic primary is 1.0~M$_\odot$. 
    The dotted line corresponds to systems 
      consisting of a 0.7-M$_\odot$ magnetic white dwarf 
      and a 0.1-M$_\odot$ non-magnetic white dwarf, and 
      the dashed line corresponds to systems 
      consisting of a 1.3-M$_\odot$ magnetic white dwarf 
      and a 0.1-M$_\odot$ non-magnetic white dwarf. 
    The magnetic moment of the primary is fixed 
      to be $10^{32}$~G~cm$^3$.       }
\end{figure}       

\begin{figure} 
\vspace*{0.2cm}  
\begin{center}
  \epsfxsize=8cm \epsfbox{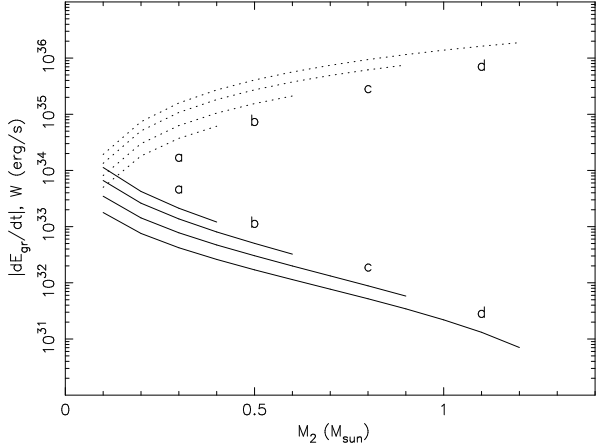}  
\end{center}
\caption{The total electrical power dissipation (solid lines) 
       and the power of gravitation radiation  
       emitted from the binary (dotted lines)  
       as a function of mass of the non-magnetic white dwarf.   
    The orbital period is fixed to be 9.5~min.  
    Curves a, b, c and d correspond to 
       systems with a 0.5, 0.7, 1.0, 
       and 1.3-M$_\odot$ magnetic white dwarf.  
    The magnetic moments of the white dwarfs are $10^{32}$~G~cm$^3$. 
    The spin of the magnetic white dwarf is 
       1 part in 1000 deviated from synchronous rotation with the orbit, 
       i.e.\ $(1-\alpha) = 0.001$. } 
\end{figure}        

If the degree of spin-orbit asynchronism $(1-\alpha)$ is specified, 
   the total power dissipation of the binary 
   can be calculated directly  
   using equations (1), (2), (4) and (5). 
We show in Figure 3 
   the total power dissipation $W$ 
   as a function of the orbital period $P$,  
   for systems with different primary and secondary white-dwarf masses 
   $M_1$ and $M_2$.  
The general trend of the dependence of $W$ on the system parameters 
   is as follows:     
  (i) the power dissipation $W$ is larger 
      for larger degree of spin-orbit asynchronism; 
  (ii) $W$ increases when $P$ decreases; 
  (iii) the larger is the radius of the non-magnetic white dwarf $R_2$, 
      the larger $W$ will be; and 
  (iv) $W$ decreases slightly 
      when $M_1$ increases.   
Moreover, provided that there is a slight spin-orbit asynchronism 
   (1 part in 1000 is sufficient) 
   and that the asynchronously rotating white dwarf 
   has a moderate magnetic moment ($\sim 10^{32}$~G~cm$^3$),        
   the electrical power dissipation of the systems 
   can reach the solar luminosity 
   $L_\odot$ ($3.9 \times 10^{33}$~erg~s$^{-1}$),   
   when the orbital period is short enough (\ltae 10~min). 
Clearly, the electrical power generated by these close binaries 
   is substantially larger than  
   the expected intrinsic luminosities of the white-dwarf components, 
   which is $\ll L_\odot$. 

It is worth noting that 
   most of the luminosity originates 
   from the two hot spots near the magnetic poles 
   at the surface of the magnetic white-dwarf.          
If the footpoints of the current-carrying fieldlines remain fixed  
   with respect to the secondary, 
   as in the Jupiter-Io system (Clarke et al.\ 1996), 
   asynchronous rotation will not manifest itself 
   in orbital-period changes, 
   but will possibly be visible through changes in the spot latitudes
   if the dipole is not aligned with the spin axis.    
   
\subsection{Spin-orbit coupling}  

The consequence of the energy dissipation 
   is the creation of a Lorentz torque,  
   which tends to synchronise the spin of the magnetic white dwarf 
   with the spin of the non-magnetic white dwarf,  
   and hence with the orbital rotation.   
Thus, through spin-orbit coupling, 
   energy and angular momentum can also be exchanged 
   between orbit and the white dwarfs. 
If the secondary white dwarf is tidally locked to the orbit,  
   as assumed in our model, 
   the evolution of the spin 
   of the magnetic primary white dwarf $\omega_1$,  
   the orbital rotation $\omega_o$, 
   and the degree of synchronisation 
   is governed by the equations:   
\begin{eqnarray}   
    {{{\dot \omega}_1}\over {\omega_1}} & = & 
     {W \over {\alpha (1-\alpha) I_1 \omega_o^2}} \ ; \\ 
    {{{\dot \omega}_o}\over {\omega_o}} & = & 
     {1 \over g(\omega_o)} 
        \biggl[{\dot E}_{\rm ex} 
      - {W \over{(1-\alpha)}} \biggr] \ ;  \\    
         {{\dot \alpha}\over \alpha} & = & 
       -{1 \over g(\omega_o)} 
        \biggl[{\dot E}_{\rm ex} 
      - {W \over{(1-\alpha)}} 
     \biggl(1+ {{g(\omega_o)}\over 
            {\alpha I_1 \omega_o^2}} \biggr) \biggr]    
   \end{eqnarray} 
(see Appendices C, D and E), where  
\begin{eqnarray}  
    g(\omega_o) & = &  -{1\over 3} 
      \biggl[{{q^3}\over {(1+q)}}G^2M_1^5\omega_o^2\biggr]^{1/3} 
        \nonumber \\ 
   &  & \hspace*{0.5cm} \times  
     \biggl[1 - {6 \over 5}(1+q)f(\omega_o) \biggr]   \ .    
\end{eqnarray}  
 
The term ${\dot E}_{\rm ex}$ is the energy loss 
   due to external process (such as gravitational radiation).  
The factor $f(\omega_o)$ 
   can be expressed in terms of orbital parameters, 
   which is simply $(R_2/a)^2$.  
Clearly, the spin-orbit evolution is jointly determined by 
   the rate of energy loss 
   due to gravitation radiation ${\dot E}_{\rm gr}$  
   and the electrical power dissipation $W$.   
   
For short-period binaries    
   gravitational radiation is an efficient process 
   to extract energy from the binary orbit 
   (Pacz$\acute {\rm n}$yski 1967; Faulkner 1971).  
As an illustration we show in Figure 4 
   the electric power dissipation $W$ 
   and the rate of energy loss from the orbit 
   due to gravitation radiation ${\dot E}_{\rm gr}$ 
   for binaries with an orbital period $<10$~min
   and spin-orbital asynchronism of 1 part in 1000. 
The power loss due to gravitational radiation is generally larger than 
   the electrical power dissipation, 
   except for systems with a very low-mass non-magnetic white dwarf.    
  
As $W \propto (1-\alpha)^2 > 0 $ (from Equations 1 and 2),   
   energy is extracted from the spin of the magnetic white dwarf, 
   if it is a fast spinner $\alpha > 1$ (Equation 9).  
When $\alpha < 1$, energy is injected 
   to spin up the magnetic white dwarf. 
Because both ${\dot E}_{\rm gr}$ and $g(\omega_o)$ are negative,  
   when energy is extracted from the orbit 
   to spin up the magnetic white dwarf,   
   the orbit shrinks and the orbital angular speed increases.   
For white-dwarf pairs consisting 
   of a 1.0-M$_\odot$ magnetic primary and a 0.1-M$_\odot$ secondary 
   with a 10-min orbit and a spin-orbital asynchronism 
   of 1 part in 1000, 
   the typical evolutionary timescale for $\alpha$ 
   is about $10^3$~years. 
As the synchronisation timescale due to unipolar induction  
   is significantly shorter than 
   the timescale of the gravitational-radiation power loss, 
   which is about $10^6$ years for these systems, 
   these systems are expected to be rare 
   in comparison with `ordinary' white-dwarf pairs.   

%Systems of this type may be driven to coalescence, 
%   because of the accelerated shrinkage of the orbit 
%   due to additional orbital energy loss by resistive dissipation.  
%This will shorten the mass-transfer phase 
%   usually expected in other close binaries, 
%   which most likely leads to mass loss from the binary 
%   (e.g.\ through novae). 
%If the magnetic and non-magnetic white-dwarf pairs 
%   with total masses exceeding the Chandrasekhar limit, 
%   they may be the progenitors of type Ia supernovae.    

\subsection{Effects of the induced magnetic field} 
   
In the unipolar-inductor model that we consider above, 
  the magnetic field generated by the currents between the white dwarfs 
  has been neglected.  
Because of this induced magnetic field,  
  the magnetic-field configuration is no longer dipolar 
  as for the assumed intrinsic field of the primary white dwarf.  
As a result, the currents and the magnetic field lines 
  are not aligned, 
  and there will an additional drag 
  to the motion of the secondary white dwarf.    
This plasma-inertia effect can be non-negligible. 

The importance of plasma-inertia effects in the Jupiter-Io system  
  was investigated by Drell, Foley \& Ruderman (1965) 
  in the linear approximation 
  and by Neubauer (1980) using a more appropriate non-linear treatment. 
These studies showed that the non-alignment of the currents and magnetic field 
  is associated with the Alfv$\acute {\rm e}$n waves 
  --- called Alfv$\acute {\rm e}$n wings --- 
  standing in the frame of Io, 
  causing dissipation in the polar regions of Jupiter as well as in Io.  

Here in this white-dwarf pair, 
  the standing Alfv$\acute {\rm e}$n waves are 
  in the frame of the non-magnetic secondary white dwarf.  
The significance of the plasma-inertia effects can be expressed by 
  the Alfv$\acute {\rm e}$n Mach number 
\begin{equation}  
  M_{\rm A} \ = \ {{\delta v} \over v_{\rm A}} \ , 
\end{equation}   
  where $\delta v$ is velocity of secondary white dwarf 
  related to the magnetosphere of the primary magnetic white dwarf, 
  given by  
  $\delta v \sim (1-\alpha)\pi a /P \approx 10^9~(1-\alpha) $~cm~s$^{-1}$, 
  and $v_{\rm A}$ is the Alfv$\acute {\rm e}$n speed:   
\begin{equation} 
 v_{\rm A} \ = \ 1.5 \times 10^9 \bigg({B \over {1~{\rm kG}}}\bigg)
  \bigg({{n_{\rm e}} \over {10^{10} {\rm cm}^{-3}}}\bigg)^{-1/2} 
  \ {\rm cm~s}^{-1} \ , 
\end{equation} 
  ($n_{\rm e}$ is the electron number density in the plasma).  
For the parameters of our interest, $M_{\rm A} \ll 1$ is generally satisfied, 
  and the plasma-inertia effects can be considered as a perturbation.  
The results that we have obtained are therefore generally valid.   

The large currents in the system can also cause significant distortion 
  of the magnetic field near the primary magnetic white dwarf. 
The geometry of the energy-dissipation region 
  and the electrical circuit of the system 
  are therefore not as simple as described in our idealised model. 
A particular issue is whether or not steady currents can be maintained  
  given the presence of a large ${\vec j} \times {\vec B}$ force, 
  when the current and the magnetic field are not aligned.  
This could eventually lead to the breakdown of the circuit 
  and quench the operation of the unipolar induction.   
Nevertheless, if the unipolar induction does operate,  
  in spite of the non-alignment of current and field,   
  the energy and angular-momentum budget 
  should be similar to that of the unipolar-inductor model 
  that we have presented above. 
(Detail investigation of the effects 
  due to severe field distortion is beyond the scope of this paper, 
  and will be left for future study.) 

\section{RX J1914+24: a possible unipolar inductor?} 

We propose that the short-period binary system RX J1914+24 
   is electrically powered.  
This system emits soft X-rays which are modulated on a period of 9.5 min,
   and whose folded light curve is 
   that expected from emission originating from one or two small spots.  
The optical/infra-red flux is also modulated only on the 9.5-min period 
   but maximum light occurs $\sim 0.4$ orbital cycles 
   earlier than the soft X-rays (Ramsay et al.\ 2000b).  

It has been argued that the system 
   is a magnetic white dwarf-white dwarf binary, 
   which is synchronised 
   by the strong magnetic interaction between the white dwarfs 
   (Cropper et al.\ 1998).  
An accretion scenario has been proposed (Ramsay et al.\ 2000a), 
   in which the optical flux is due to irradiation 
   by the accreting material, 
   and which can largely account for the observations. 
However, to explain the lack of optical polarisation (Ramsay et al.\ 2000a), 
   the magnetic-field strength on the primary 
   must either be too high or too low to be detected 
   in the optical/infra-red.  
Also, a major concern is that the accretion model is unable 
   to explain the observed lack of strong emission lines 
   (Ramsay et al.\ 2000b)     
   which result from heating of the accretion stream 
   in mass-transfer systems.  

If the secondary in RX~J1914+24 fills its Roche lobe 
   (Cropper et al.\ 1998), 
   $M_2 \approx 0.1 {\rm M}_\odot$.  
A Roche-lobe-filling star is more easily tidally locked 
   into synchronous rotation with the orbit. 
In the unipolar-inductor model proposed above,  
   the secondary is allowed to lie within its Roche lobe,         
   and so this mass is a lower limit.  
The mass of the primary is, however, unconstrained.
Based on {\it ROSAT} and {\it ASCA} measurements 
   and an estimated distance of 200-500 pc, 
   the deduced luminosity of RX J1914+24 is in the range 
   $4 \times 10^{33}$ to $1 \times 10^{35}~{\rm erg~s}^{-1}$
   (Ramsay et al.\ 2000b). 
Provided that the spin-orbit asynchronism 
   of the magnetic white dwarf is about 1 part in 1000 or larger, 
   the resistive heating is sufficient to power these luminosities 
   (See Figures 3 and 4).   
From Equation (4), 
   the resistive dissipation power is estimated 
   to be $\sim 10^{31}-10^{32}~{\rm erg~s}^{-1}$ 
   in the secondary's atmosphere.   
The result is less sensitive to other system parameters 
   such as the mass of the magnetic white dwarf.    
A comparison between the observed and predicted luminosities  
   suggests that the secondary is 0.15~M$_\odot$, 
   and thus close to filling its Roche lobe and tidally locked. 
 
The predicted area of each hotspot on the primary is 
   $\sim 8\times 10^{14}~{\rm cm}^2$ (from Appendix A).   
If we use the observed luminosity of $10^{34}~{\rm erg~s}^{-1}$ 
   (assuming a distance $d = 300$~pc), 
   we obtain a blackbody temperature of 50~eV,
   consistent with the measured value of 55~eV 
   from the {\it ROSAT} and {\it ASCA} spectral fits 
   (Ramsay et al.\ 2000b).  
This consistent area is corroborating evidence that 
   the underlying geometry of the model is valid. 

For a system 
   with a 1.0-M$_\odot$ primary, a 0.1-M$_\odot$ secondary  
   and an orbital period of 9.5~min, 
   less than 1 percent of the X-ray flux 
   emitted from the hotspots primary is intercepted 
   by the secondary. 
In equilibrium, the intercepted flux will be reradiated,  
   resulting in an increase in the luminosity 
   of the irradiated hemisphere of the secondary. 
The additional luminosity arising from irradiation heating 
   is estimated to be  
   \ltae $10^{32}~{\rm erg~s}^{-1}$. 
Thus, the power of irradiative heating of the secondary 
   is similar to the power of electrical heating.  
Given the fact that the two heating processes have similar efficiency, 
   one would not expect a large temperature  difference  
   in the irradiated and the unirradiated hemisphere of the secondary.  
This is consistent with the observation that   
   the amplitude of variation in the optical luminosity   
   in RX~J1914+24 is small (Ramsay et al.\ 2000a).  
Moreover, as the power of irradiative heating 
   is of the same order of the energy flux 
   from below the atmosphere due to resistive heating,  
   the presence of moderate heat conduction is sufficient 
   to maintain an approximate isothermal layer 
   down to unit optical depth. 
Because of the absence of an temperature inversion layer,  
   the secondary will not show emission lines in its spectrum: 
   this is consistent with the lack of prominent emission lines 
   in the optical spectrum (Ramsay et al.\ 2000b). 

\section{Summary}

We propose that short-period 
   magnetic and non-magnetic white-dwarf pairs 
   with short orbital periods ($\sim 10$~min) 
   are efficient cosmic unipolar inductors.        
Provided that the spin of the magnetic component and the orbit 
   are not in perfect synchronism, 
   a large e.m.f.\ can be produced across the non-magnetic white dwarf.   
The resistive dissipation in the white dwarfs is sufficient to power 
   luminosities significantly above solar values;  
   most power is dissipated at the hot spots 
   on the surface of the magnetic white dwarfs, 
   which are footpoints of the field lines connecting the two stars.  
Electrical power is therefore an alternative luminosity source, 
   following on from nuclear fusion and accretion.   

The X-ray source RX~J1914+24 
   is a candidate unipolar inductors 
   consisting of a magnetic and non-magnetic white-dwarf pair.  
The two small X-ray spots on the magnetic white dwarf 
   predicted by the unipolar-inductor model 
   are compatible with the X-ray light curve of RX~J1914+24.  
The luminosity and temperature predicted by the model is also 
   in agreement with the observed values derived 
   from fits to the X-ray spectra.  
The model also explains the variation in the optical/infra-red luminosity, 
   and the detection of only a single period. 
The variations in the long term X-ray intensity 
   can be attributed to variations in the current flow.    
The two main inadequacies of the current accretion model
   (Ramsay et al.\ 2000a) 
   --- the lack of any polarised flux 
   and the lack of any detectable line emission (Ramsay et al.\ 2000b) 
   are naturally explained. 

\section*{Acknowledgments}  
KW acknowledges the support from the Australian Research Council 
  through an ARC Australian Research Fellowship.  
We thank Jan Kuijpers, Jianke Li, Qinghuan Luo and Mark Wardle 
  for discussions and critical reading of the manuscript. 
We also thank Jan Kuijpers and Don Melrose 
  for pointing to our attention of effects 
  due to the Alfv$\acute {\rm e}$n wings.

\section*{Appendix A: Geometry of the hot spots}   

Consider a system consisting of a magnetic and a non-magnetic 
  white dwarf in close orbit with a separation $a$. 
The magnetic white dwarf has a dipolar field, 
  with the family of field lines given by 
  $r = {\cal C}~{\sin}^2 \theta$, 
  where ${\cal C}$ is a constant labeling different field lines. 
Thus, the geometry of the pole is determined by 
  the field lines that connected the two white dwarfs, 
  and it is specified by the parameter 
  $\tilde a$ and $\tilde b$, where 
\begin{eqnarray}  
   {\tilde a} & = &  
       {{R_1} \over 2} \sin \theta \Delta \phi  \nonumber \\ 
        & = &  \biggl({{R_1^3}\over a}\biggr)^{1/2}
            \tan^{-1} \biggl({{R_2} \over a} \biggr) \ ,  \\ 
   {\tilde b} & = &   {{R_1} \over 2}  \Delta \theta \nonumber \\  
   & = &    {{R_1} \over 2}
     \big( \theta^{-} - \theta^{+} \big) \nonumber \\ 
      & = &  {{R_1} \over 2}
       \bigg\{ \sin^{-1} {\sqrt{{R_1}\over {a-R_2}}} 
       - \sin^{-1} {\sqrt{{R_1}\over {a+R_2}}}  \bigg\}  
\end{eqnarray}   
(see Figure 5, top and second panels).    

If the radius of the stars $R_2$ and $R_1$ 
   are significantly smaller than the orbital separation $a$, 
   then ${\tilde a} \approx R_2  (R_1/ a)^{3/2}$ and     
   ${\tilde b} \approx (R_2/2)  (R_1/ a)^{3/2}$. 
It follows that ${\tilde b}/ {\tilde a}$ is approximately 0.5 
   for a wide range of orbital parameters.   
Moreover, the fractional area of the dissipative region 
   on the white-dwarf surface $f<({\tilde a}{\tilde b}/R_1^2) \ll 1$. 

\section*{Appendix B: Resistive dissipation in the two white dwarfs}  

\begin{figure} 
\vspace*{0.2cm}  
\begin{center}
  \epsfxsize=7.2cm \epsfbox{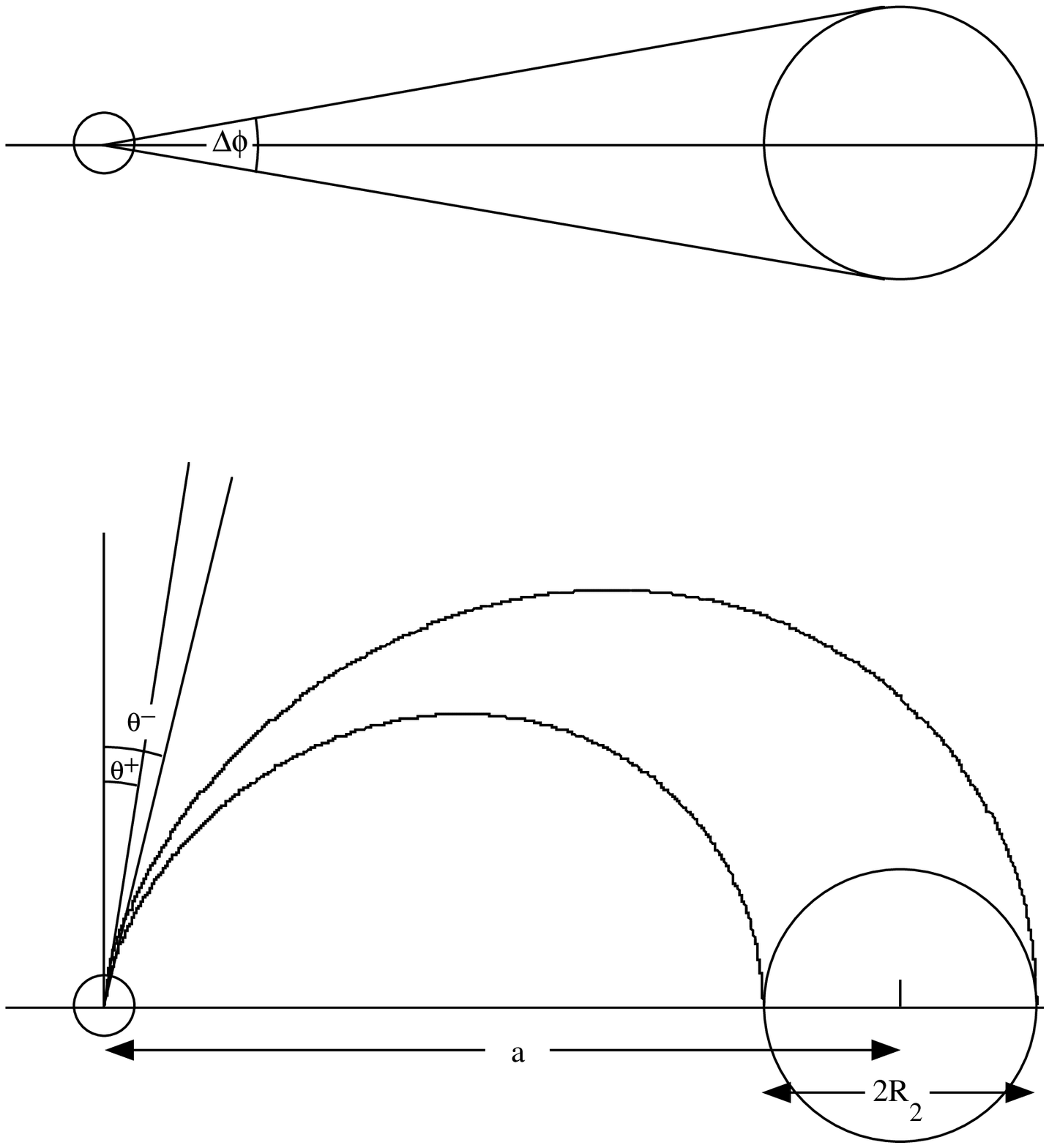}   \\ 
  \epsfxsize=5cm \epsfbox{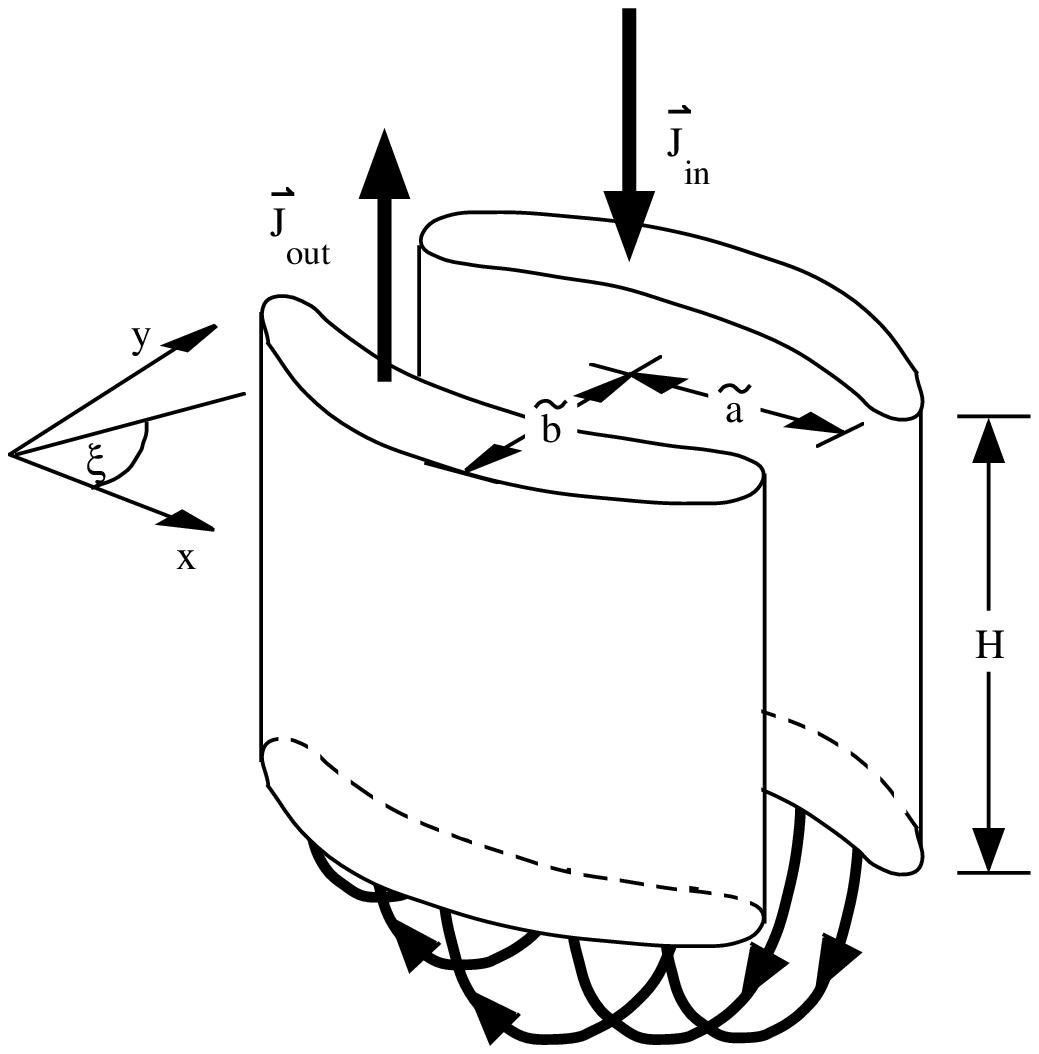}   \\ 
  \epsfxsize=5cm \epsfbox{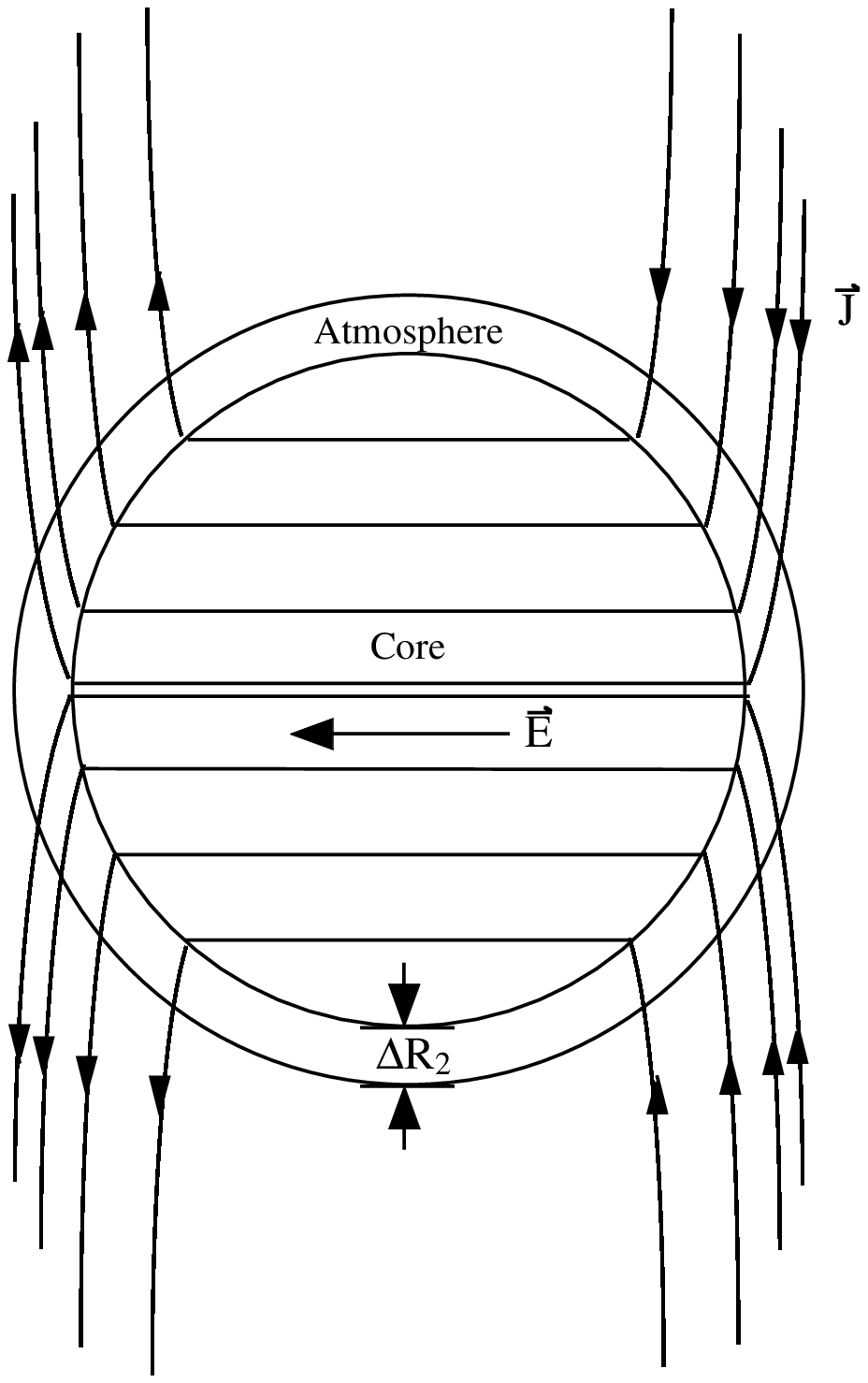}      
\end{center}   
\caption{(Top and second panels) 
  The top and side elevation views of the system   
      showing how maximum height and width of the hot spot 
      (2$\tilde b$ and 2$\tilde a$ respectively)  
      at the surface of the magnetic white dwarf  
      are determined.   
  (Third penal) Schematic illustration of the current sheets 
      in the atmosphere of the magnetic primary white dwarf. 
  At depth $H$ below the white-dwarf, 
     collisions are efficient enough 
     to allow the current to cross the magnetic-field lines. 
  Bottom panel) Schematic illustration of the current flow 
     in the non-magnetic secondary white dwarf. } 
\end{figure}      

For the non-magnetic white dwarf, 
   the conductivity of the core $\sigma_{\rm co}$  
   is much larger than the conductivity of the atmosphere $\sigma_2$
   (i.e.\ $\sigma_{\rm co} \gg \sigma_2$), 
   so that most resistivity dissipation occurs at the atmospheric layer 
   (See Figure 5, bottom panel).   
The rate of energy dissipation is therefore 
\begin{eqnarray}  
   W_2 & = & \int_{R_2 - \Delta R_2}^{R_2} d^3x \  
            \big({{\vec J}_2 \cdot {\vec E}}\big)   
                \nonumber \\ 
        & = & \int_{R_2 - \Delta R_2}^{R_2} d^3x \   
               {1 \over \sigma_2} | {\vec J}_2 |^2 \ . 
\end{eqnarray} 
If the current density $\vec J$ is roughly homogeneous, then   
\begin{eqnarray}  
   W_2 & = & 4 \pi R_2^2 {\Delta R_2} {J_2^2 \over \sigma_2}  
                \nonumber \\ 
       & \approx & {4 \over \pi} {I^2 \over {\sigma_2 R_2}} 
                  \biggl({{\Delta R_2}\over {R_2}} \biggr) \ ,  
\end{eqnarray} 
   where $I = \pi R_2^2 J_2$ is the current flowing across the star.  

For the magnetic white dwarf, the dissipation at each polar spot is 
\begin{eqnarray}  
   W_1^{(1)} & = &  
        \int_{\rm spot} d^3x\  \big({{\vec J}_1 \cdot {\vec E}}\big)  \ .  
\end{eqnarray} 
 
It is not trivial to determine   
   the functional form of the current density ${\vec J}_1({\vec x})$ 
   at the pole.  
For simplicity, we follow Li, Wickramasinghe and Ferrario (1998) 
   and consider an approximation that the current density is non-zero 
   in two arc-like regions. 
The thickness of the arc is $\Delta d$, 
   and the length and width of the region are 
   $2{\tilde a} = R_1 \sin \theta \Delta \phi$ and 
   $2{\tilde b} = R_1 \Delta \theta$, 
   where $\theta$ is the colatitude of the centre of the hot spot.   
The current density is given by $J_1 = J_{1o} \sin \xi$ 
   (Figure 5, third panel).   
We assume that at a depth $H (\ll R_1)$ the Pendersen conductivity 
   (see Alfv$\acute {\rm e}$n \& F$\ddot {\rm a}$lthammar 1963) 
   becomes sufficient large     
   to allow the current to cross the field lines and close the circuits.    
 
The current flowing through the spot is therefore 
\begin{eqnarray}  
    I & = & \int_{\rm spot} dxdy \ J_1  \nonumber \\ 
      & = & J_{\rm 1o} (4 {\tilde a} {\Delta d} )  
         \int_0^{\pi/2} d\xi  \  {\cal Q}_1(\xi; e)    \  , 
\end{eqnarray}   
   and the resistive dissipation is 
\begin{eqnarray}  
   W_1^{(1)} & = &  {{J_{1o}^2}\over {\sigma_1}} (4 {\tilde a} H \Delta d)  
         \int_0^{\pi/2} d\xi \  {\cal Q}_2(\xi; e)   \ .  
\end{eqnarray}   
The function  ${\cal Q}_n$ is given by 
\begin{eqnarray} 
 {\cal Q}_n(\xi; e) & =  & 
    {\sin^n \xi} \sqrt{1 - e^2 {\cos^2\xi}} \ ,  
\end{eqnarray}   
 where $e = \sqrt{ 1 - ({\tilde b} / {\tilde a})^2}$.  

Using the expression for $\tilde a$ in Appendix A, 
  we obtain the rate of total dissipation at the magnetic white dwarf:   
\begin{eqnarray}  
  W_1 & = & 2 W_1^{(1)}  \nonumber \\ 
      & \approx & {1\over 2} {I^2 \over {\sigma_1}}  
      \biggl({{H}\over {\Delta d}} \biggr)
      \biggl( {a \over R_1} \biggr)^{3/2} {{\cal J}(e) \over {R_2}}  \ , 
\end{eqnarray} 
  where 
\begin{eqnarray}
    {\cal J}(e) & = & 
      {{\int_0^{\pi/2} d\xi \  {\cal Q}_2(\xi; e)}
         \over {\bigl[\int_0^{\pi/2} d\xi  
            \  {\cal Q}_1(\xi; e)  \bigr]^2}} \nonumber \\ 
      & = & \pi ~\bigg\{ 1 - \sum_{n=1}^{\infty} 
            \biggl[{{(2n-1)!!}\over {(2n)!!}} \biggr]^2  
            {{e^{2n}} \over {(n+1)(2n-1)}}  \bigg\}   \nonumber \\   
   & & \times 
     \biggl[\sqrt{1-e^2}+{1\over e}\sin^{-1}e \biggr]^{-2}   \ .  
\end{eqnarray}   
The ratio of the dissipation rates at the two white dwarfs is therefore 
\begin{eqnarray}  
  {{W_1} \over {W_2}} & = & {\pi \over 8} {\cal J}(e) 
        \biggl({{\sigma_2} \over {\sigma_1}}\biggr)  
        \biggl({{H}\over {\Delta d}}\biggr) 
        \biggl({{R_2}\over {\Delta R_2}}\biggr) 
        \biggl({{a}\over {R_1}}\biggr)^{3/2}  \  . 
\end{eqnarray} 
The exact values of $H$ and $\Delta d$ are uncertain. 
Li, Wickramasinghe \& Ferrario (1998) argued that 
   roughly $H \sim \Delta d$.    
As $\sigma_2 \sim \sigma_1$, 
   $\pi {\cal J}(e)/ 8  \sim {\cal O}(1)$ and 
   $({R_2}/{\Delta R_2}) ({a}/{R_1})^{3/2} \gg 1$,   
   most of the electric energy will be dissipated  
   at the two hot polar spots of the magnetic white dwarf.  
                     
\section*{Appendix C: Energy conservation}  
                         
Consider a system with two gravitationally bound, rotating objects,  
  with moments of inertia $I_1$ and $I_2$, in circular motion 
  around each other. 
Let the angular speeds of their rotation be 
  $\omega_1$ and $\omega_2$ respectively.   
The total mechanical energy of the system is 
\begin{equation} 
   E \ = \  {1 \over 2} I_1  \omega_1^2  
            + {1 \over 2} I_2  \omega_2^2   + (T+V) \  . 
\end{equation} 
$T$ and $V$ are the kinetic and potential energy of the orbit, 
   which are related by 
\begin{eqnarray}  
   T & = & {1\over 2} I_o  \omega_o^2 \nonumber   \\  
     & = & {1\over 2} {{G M_1^2 q} \over a }  \\ 
     & = & - {1\over 2} V \nonumber  \ , 
\end{eqnarray}    
   where the effective moment of inertia of the orbit 
   $I_o = M_1 a^2 [q/(1+q)]$ and    
   the orbital angular speed                  
   $\omega_o = [G M_1 (1+q)/a^3]^{1/2}$.
   
If the spin of the second object is in perfect synchronism with the orbit 
   (i.e.\ $\omega_2 = \omega_o$), then 
\begin{eqnarray}  
   E & = &  {1 \over 2} I_1  \omega_1^2    
        - {1 \over 2} \biggl[ M_1 \biggl( {q \over {1+q}} \biggr) a^2 
                - {2 \over 5} M_1 q R_2^2 \biggr] \omega_o^2 \nonumber \\ 
      & = & {1 \over 2} I_1  \omega_1^2  
        - {1 \over 2} {{GM_1^2 q} \over {a^3}} 
          \biggl[ a^2 - {2 \over 5} (1+q) R_2^2 \biggr]  \ .  
\end{eqnarray}   
(Here, we have assumed that the objects are spherical, 
   so that their moments of inertia is 
   2/5 of their mass times the square of their radius.)  
It follows that 
\begin{eqnarray}   
   {\dot E} & = & I_1  \omega_1 {\dot \omega}_1  
      - {1 \over 2} G M_1^2 q  {d \over {dt}}  \biggl[ {1 \over a}  
      - {2 \over 5} (1+q) {{R_2^2} \over {a^3}} \biggr] \nonumber \\ 
   & = & I_1 \omega_1 {\dot \omega}_1 
      +{1 \over 2} {{G M_1^2 q}\over a}  \biggl[ 1   
      - {6 \over 5} (1+q) \biggl({{R_2} \over a}\biggr)^2 \biggr] 
        \biggl( {{\dot a}\over a}\biggr)  \nonumber \\ 
   & = &  I_1 \omega_1 {\dot \omega}_1 
      -{1 \over 2} {I_o \omega_o^2}  \biggl[ 1   
      - {6 \over 5} (1+q) f(\omega_o)  \biggr] 
        \biggl({2\over 3}{{{\dot \omega}_o}\over {\omega_o}}\biggr) \  ,  
\end{eqnarray}   
  where $f(\omega_o) = \{ {R_2^3 \omega_o^2}/ [GM_1(1+q)]\}^{2/3}$. 
If the second object is close to filling its Roche-lobe, 
  then $R_2 \approx a \lambda [q/(1+q)]^{1/3}$, and  
\begin{equation}   
   {\dot E}\ \approx \  I_1  \omega_1 {\dot \omega}_1    
      +{1 \over 2} {I_o \omega_o^2}  \biggl[ 1   
      - {6 \over 5} \lambda^2 (1+q)^{1/3}q^{2/3}  \biggr] 
        \biggl({{\dot a}\over a}\biggr) \  ,  
\end{equation}   
  where $\lambda = 0.462$.  
Clearly, when $q < 0.7$, $\lambda^2 (1+q)^{1/3}q^{2/3} < 5/6$, 
  and hence the square bracketed term is positive.   
                               
\section*{Appendix D: Work done by the dissipative torque} 
 
The torque that accelerates/decelerates the spin of the object 1 is 
\begin{eqnarray} 
   {\vec \tau} & \equiv & I_1 \ddt {\vec \omega}_1 \  , 
\end{eqnarray} 
   and the torque that changes the spin of object 2 
   and the orbital rotation is 
\begin{eqnarray}  
  {\vec \tau}_{\rm ex} - {\vec \tau} & = & I_o \ddt {\vec \omega}_o 
           + I_2 \ddt {\vec \omega}_2  + {\vec \omega}_o \ddt {I_o} \ , 
\end{eqnarray} 
  where $\tau_{\rm ex}$ is an external torque.    

From the definition of the synchronisation 
   $\alpha \equiv \omega_1/\omega_o$, 
we have 
\begin{eqnarray}  
  {{\dot \alpha}\over {\alpha}} & = & 
             {{{\dot \omega}_1}\over {{\omega}_1}} 
           - {{{\dot \omega}_o}\over {{\omega}_o}}  \ . 
\end{eqnarray}  
The perfect-synchronism condition for object 2 and the orbit implies that 
\begin{eqnarray}  
  \tau_{\rm ex} - \tau & = & (I_o+I_2) {\dot \omega}_o 
           + \omega_o {\dot I}_o \ .  
\end{eqnarray}    
Thus, the timescale for the change in the orbital angular speed is 
\begin{equation}  
   {{{\dot \omega}_o} \over {\omega_o}} \ = \  
      K \biggl[ {\tau_{\rm ex} \over {(I_o+I_2)\omega_o}} 
         -  \biggl({{\alpha I_1}\over {I_o +I_2}}\biggr) 
         {{\dot \alpha} \over \alpha} 
         - {{{\dot I}_o} \over {(I_o+I_2)}}  \biggr]  \ , 
\end{equation}  
and the timescale of spin evolution of object 1 is 
\begin{equation}  
   {{{\dot \omega}_1} \over {\omega_1}} \ = \  
      K \biggl[ {\tau_{\rm ex} \over {(I_o+I_2)\omega_o}} 
         +
         {{\dot \alpha} \over \alpha} 
         - {{{\dot I}_o} \over {(I_o+I_2)}}  \biggr]  \ , 
\end{equation}  
  where $K = [1 + \alpha I_1/(I_o +I_2)]^{-1}$.    
The first term in the square of above expression is 
  due to the synchronisation process, 
  the second term is the contribution of the external torque, 
  and the last term is caused by the readjustment of the orbital separation. 

The work done by the (dissipative) synchronisation torque is 
\begin{eqnarray}   
  W & = &  |{\vec \tau} \cdot({\vec \omega}_o -{\vec \omega}_1)| 
         \nonumber \\       
    & = &  | I_1 {\dot \omega}_1 \omega_o (1- \alpha) | \nonumber \\ 
    & = &  \bigg\vert I_1 \omega_o^2 \alpha (1- \alpha)  
       \biggl[ {{\dot \alpha}\over {\alpha}} 
      + {{{\dot \omega}_o}\over{{\omega}_o}} \biggr] \bigg\vert \ .  
\end{eqnarray}    
It is straightforward to show that 
   the work done can also be expressed as follows: 
\begin{eqnarray} 
  W  & = &  \bigg\vert I_1 \omega_o^2 \alpha (1- \alpha) 
     \biggl[ 1 + {{\alpha I_1}\over {I_o +I_2}} \biggr]^{-1}
      \nonumber \\      
   &  & \hspace*{0.5cm} \times 
    \biggl[ {\tau_{\rm ex} \over {(I_o+I_2)\omega_o}} 
         + {{\dot \alpha} \over \alpha} 
         - {{{\dot I}_o} \over {(I_o+I_2)}}  \biggr] \bigg\vert  \ . 
\end{eqnarray}   

\section*{Appendix E: Spin-Orbit evolution}  

Suppose the energy loss is represented 
  by a dissipative term due to synchronisation torque 
    $\dot E_{\rm diss}$
  and an additional term $\dot E_{\rm ex}$
    (e.g.\ due to gravitational radiation). 
Then we have 
\begin{eqnarray}  
   {\dot E}_{\rm diss} - {\dot E}_{\rm ex} & = & 
     I_1 \omega_1 {\dot \omega}_1  + g(\omega_o) 
         \biggl({{\dot \omega_o}\over {\omega_o}}\biggr) \ ,      
\end{eqnarray}  
   where 
\begin{eqnarray} 
   g(\omega_o) & = &  -{1\over 3} 
      \biggl[{{q^3}\over {(1+q)}}G^2M_1^5\omega_o^2\biggr]^{1/3} 
        \nonumber \\ 
   &  & \hspace*{0.5cm} \times  
     \biggl[1 - {6 \over 5}(1+q)f(\omega_o) \biggr] \ .  
\end{eqnarray}  
The functional form $f(\omega_o)$ can be found in Appendix C.   
For dissipative synchronisation torque 
   proportional to $({\vec \omega}_o - {\vec \omega}_1)$, 
   we have    
\begin{eqnarray}  
   {\dot E}_{\rm diss} & = & -W  \nonumber \\ 
    & = & -(1-\alpha) I_1 \omega_o {\dot \omega}_1 \ . 
\end{eqnarray}  
It can be shown that 
   the evolution of the spin of the magnetic white dwarf 
   and the orbital rotation are determined by the following equations:  
\begin{eqnarray}    
    {{{\dot \omega}_o}\over {\omega_o}} & = & 
     {1 \over g(\omega_o)} 
        \biggl[{\dot E}_{\rm ex} 
      - {W \over{(1-\alpha)}} \biggr] \ ;  \\   
       {{\dot \alpha}\over \alpha} & = & 
       -{1 \over g(\omega_o)} 
        \biggl[{\dot E}_{\rm ex} 
      -  {W \over{(1-\alpha)}} 
     \biggl(1+ {{g(\omega_o)}\over 
            {\alpha I_1 \omega_o^2}} \biggr) \biggr] \ . 
\end{eqnarray}

\end{document}